\documentclass[11pt,a4paper]{article}
\usepackage{amssymb,amsfonts,amsmath,amsthm,euscript}
\usepackage[english]{babel}
\usepackage[utf8]{inputenc}
\usepackage{dsfont}
\usepackage{graphicx}
\usepackage{subfigure}
\usepackage{color}
\usepackage{mathrsfs}
\usepackage{enumerate}
\usepackage{graphics}
\usepackage{multirow}
\usepackage{multicol}
\usepackage{rotating}
\usepackage{placeins}
\usepackage{natbib}
\usepackage{adjustbox}
\usepackage{pdflscape}
\usepackage[final]{pdfpages}
\usepackage[normalem]{ulem}
\usepackage[hyperfootnotes=false, colorlinks=true,  linkcolor=blue, citecolor=blue,urlcolor=blue]{hyperref}
\usepackage[left=2.7cm,right=2.7cm,bottom=2.67cm,top=2.67cm]{geometry}
\allowdisplaybreaks[4]

\newcommand{\Z}{\mathds Z}
\newcommand{\R}{\mathds R}

\newcommand{\C}{\mathds C}
\newcommand{\rr}{\texttt{R }}

\newcommand{\var}{\mathop{\mathrm{Var}}}

\newcommand{\pumis}[1]{{\expandafter \xout \expandafter{#1}}}
\long\def\sfootnote[#1]#2{\begingroup%
\def\thefootnote{\fnsymbol{footnote}}\footnote[#1]{#2}\endgroup}
\def\bfootnote{\xdef\@thefnmark{}\@footnotetext}

\begin{document}
\thispagestyle{empty}
{\centering
\Large{\bf Estimation of Long-Range Dependent Models with Missing Data: to Impute or not to Impute?}\vspace{.5cm}\\
\normalsize{ {\bf Guilherme Pumi${}^{\mathrm{a,}}$\sfootnote[1]{Corresponding author. This Version: \today}\let\thefootnote\relax\footnote{\hskip-.15cm${}^\mathrm{a}$Programa de P\'os-Gradua\c c\~ao em Estat\'istica - Universidade Federal do Rio Grande do Sul.}, Gladys Choque Ulloa${}^{\mathrm{a}}$ and Taiane Schaedler Prass${}^{\mathrm{a}}$
}} \\
\let\thefootnote\relax\footnote{E-mails: guilherme.pumi@ufrgs.br (G. Pumi), gladyschoqueulloa7@gmail.com  (G.U. Choque) and taianeprass@ufrgs.br (T.S. Prass)}
\let\thefootnote\relax\footnote{ ORCIDs: 0000-0002-6256-3170 (G. Pumi); 0000-0003-3136-909X (T.S. Prass).}}

\begin{abstract}
Estimation of time series under the presence of missing data is generally a difficult and delicate problem, due to the complex manner in which it affects the model's dependence structure. In the context of long-range dependence, the problem is even more challenging and the literature regarding model estimation in the presence of missing data is very sparse. There are two basic approaches to dealing with the problem: missing data can be imputed using some plausible method; or one can use specially tailored methodologies. In this work, we review some of the methods available for both approaches and present an extensive Monte Carlo simulation study to compare their performance. We consider 35 different setups to estimate $d$ under a wide range of scenarios, contemplating percentages of missing data from as few as 10\% up to 70\% and several levels of dependence. We also provide a time benchmarking exercise comparing these methods. %To illustrate the practical relevance of our findings, we apply the intrinsic estimators to a real-world dataset of PM$_{2.5}$ concentrations from seven monitoring stations across North America, with missing data proportions ranging from 4\% to 62\%. 
\vspace{3mm}

\end{abstract}
\textbf{Statements and Declarations:} The authors declare that they have NO affiliations with or involvement in any
organization or entity with any financial interests in the subject matter or materials discussed in this manuscript.
\section{Introduction}

Long-range dependent processes have a long history and through time the subject has evolved into an essential component of time series analysis -- see the book by \cite{palma2007long}, the compilations by \cite{biblia}  and \cite{robinsonbook}, and, for an account of the history and early days' developments, the book by \cite{beran}. In this work we are interested in the class of ARFIMA$(p,d,q)$ models introduced by \cite{Hosking1981} and \cite{Granger1980}, which is one of the most applied and studied classes of long-range dependent models in the literature. There are numerous estimation procedures for the long-range dependence parameter $d$, including methods based on state-space representation, spectral density, approximations to the likelihood, wavelets, and detrended fluctuation analysis (DFA) \citep[][and references therein]{hurst1951long,geweke1983estimation,dfa,robinson1995gaussian,robinson1995log,abry1998wavelet,ChanPalma,palma2007long,fay}. Monte Carlo simulation studies comparing different estimators have also been conducted \citep{taqqu_rev, Koko, silvia, fay, rea}. Estimating $d$ in the presence of missing data, on the other hand, is a much less studied problem with sparse literature.

There are two approaches for handling the general problem of missing data in time series. The first one, imputation, is the most widely used. The basic idea is to replace missing data with plausible values, and then proceed with the analysis as if no data were missing. Of course, there are many ways to do this. One of the simplest methods is to replace missing values with the sample mean or median of the observed data, particularly in the context of stationary time series. For nonstationary time series, a missing value can be replaced with the average of previous values up to that point, or by calculating the mean locally using a sliding window approach. When there are just a few missing values corresponding to a small percentage of the total sample size, it is expected that any reasonable imputation method applied to a stationary time series should yield good results in the sense that estimated quantities should be close to those obtained if no data were missing, especially for point estimation. Imputation has the advantage of being quick and simple to implement, which adds to its appeal.

Despite their strengths, the quality of imputation-based estimation degrades rapidly as the number of missing values increases. Furthermore, almost all conventional deterministic imputation methods suffer from three main problems in the context of time series. First, variances tend to be underestimated, leading to biases in other parameters (like correlations) that depend on variances. Mean/median substitution, for example, replaces the presumably different missing values with a single value, reducing variance. It also has an effect on stationary distributions, artificially creating a point of mass in an otherwise continuous distribution. The second problem is that imputing an exogenous value into a time series changes the dependence structure in ways difficult to quantify or even understand. This is especially problematic considering that estimation in models of practical interest takes into account the time series' dependence structure. The third problem is equally serious. Standard error calculations presume that all data is accurate. The inherent uncertainty and sampling variability in the imputed values are not taken into account. As a result, standard errors and $p$-values may be distorted, leading to incorrect confidence intervals and hypothesis testing, as well as potentially misspecified models. These drawbacks generally discourage the use of simple imputation methods in the context of time series, especially when the number of missing points is considerable.

The second approach to estimating $d$ in the presence of missing data involves using estimators specifically designed for this purpose. Unlike imputation methods, these estimators inherently account for missing observations and avoid the associated pitfalls. However, such estimators remain scarce in the literature, primarily because designing them is a non-trivial problem and deriving their theoretical properties is fundamentally challenging. Consequently, the available options are often more complex to implement and computationally intensive than imputation-based solutions.

In this paper we address several important questions of a practical nature:
how high the percentage of missing data must be so that we start loosing trust on (or stop being able to compute) a given methodology? Does the strength of the dependence affect this percentage at all? Must we always use an estimator adapted or specially made for missing data, or can we use the much simpler and faster approach of imputation? Is this answer influenced by the percentage of missing and/or strength of the dependence? 

To gain insight into these  questions, we conduct a series of Monte Carlo simulation studies. We consider three different estimators specially tailored for missing data in different configurations, five of the most commonly used semiparametric estimators in the literature, paired with three different approaches to imputation, under 28 different scenarios of dependence strength and percentage of missing values. We also introduce a random imputation method especially tailored to closely mimic the original variance of the time series without introducing any potential outliers and while taking into account the time series' local dependence structure.

% To complement our simulation study, we illustrate the practical relevance of these methodologies by applying the intrinsic estimators to a real-world dataset of PM$_{2.5}$ concentrations from seven monitoring stations across North America. The stations present missing data proportions ranging from 4\% to 62\% and represent diverse environmental contexts, including rural parks, urban-industrial sites, and remote locations. This empirical application allows us to assess the behavior of the estimators under realistic conditions and to examine whether the patterns observed in our Monte Carlo simulations translate to practical settings.

\section{Framework}
We say that $\{Y_t\}_{t\in\Z}$ is an ARFIMA$(p,d,q)$ if it satisfies the difference equation
\begin{equation}\label{eq:Palma}
     \phi(L)Y_t=\theta(L)(1-L)^{-d} \varepsilon_t,
\end{equation}
 where $\{\varepsilon_t\}$ is a zero mean white noise with $\sigma_\varepsilon^2:=\var(\varepsilon_t)<\infty$, $L$ is the backward shift operator, $\phi(L):=1-\phi_1L-\cdots-\phi_pL^p$ and $\theta(L):=1+\theta_1L+\cdots+\theta_qL^q$ are the AR and MA operators, respectively;  $(1-L)^{-d}$ is a fractional differencing operator defined by the binomial expansion
\[(1-L)^{-d}=\sum_{j=0}^{\infty}\eta_j L^j, \qquad \mbox{ with } \quad   \eta_j:=\frac{\Gamma(j+d)}{\Gamma(j+1)\Gamma(d)},\]
for $-1<d<\frac{1}{2}$. It is usual to assume that $\phi$ and $\theta$ have no common roots and all roots of $\phi$ lie outside the unitary circle $\{z\in\C:|z|= 1\}$. Under these conditions, there exists a unique stationary solution for \eqref{eq:Palma} with autocovariance satisfying
\begin{equation}\label{gcovdec}
    \gamma(h)\sim \kappa_d|h|^{2d-1},\qquad \mbox{with}\quad \kappa_{d}:=\sigma_\varepsilon^2\bigg|\frac{\theta(1)}{\phi(1)}\bigg|^2\frac{\Gamma(1-2d)}{\Gamma(1-d)\Gamma(d)},\quad \mbox{as $h\rightarrow \infty$}.
\end{equation}
 For $0<d<1/2$, the covariance in \eqref{gcovdec} decays at a hyperbolic rate so that the autocorrelation is not absolutely summable.

In this work, our main interest is estimating the long-range parameter $d$ under the presence of missing values. We impose no restriction on the data missing mechanism, as all methods discussed here are indifferent to it. We also assume, without loss of generality, that the time series' first and last values are never missing. We consider two approaches for dealing with missing values.  The first is by employing a method specifically designed to estimate $d$ without the need for missing data completion. Estimators in this class will be called \emph{intrinsic estimators}. The second approach involves using an estimator that requires the time series to have no missing data. We proceed by imputing the missing values and then applying the estimator as if no values were missing. In this work, we shall consider five of the most commonly applied estimators for $d$, besides the intrinsic estimators, which can also be used when the time series is complete, and three methods for data imputation. In total, 35 different estimation procedures will be compared.  In the following sections, we will briefly review each procedure.

\subsection{Intrinsic methods}\label{native}

To the best of our knowledge, the earliest intrinsic estimator for $d$ in the presence of missing values is \cite{ChanPalma}, which introduces a state-space representation of an ARFIMA model and a modification of the Kalman filter to approximate the likelihood function in the presence of missing data. Although the methodology performed reasonably well in the authors' simulations, it suffers from being very slow, especially when compared to other alternatives, and as a result, it was not included in our simulation study.

In this work, we consider two different intrinsic estimators for $d$: two semiparametric (first and second generation) wavelet-based estimators proposed by \cite{knight2017wavelet} and \cite{craigmile}. They rely on a relationship between the undecimated wavelet variance and the parameter $d$. The former estimates the wavelet variance using wavelet lifting while the latter uses a specially designed estimator. In the next two sections, we present details regarding these intrinsic estimators.

\subsubsection{Knight et al. (2017)'s LoMPE estimator} \label{lompe}
A first generation wavelet-based estimator for the long memory parameter $d$ is presented in \cite{knight2017wavelet}. The method is based on a multiscale lifting transformation known as LOCAAT (lifting one coefficient at a time) proposed by \cite{jansen2001scattered}. The LoMPE's idea is to apply the LOCAAT method to obtain a collection of lifting coefficients, which, after suitable normalization, are used to estimate the wavelet's coefficient variance. To estimate $d$, a regression approach similar to \eqref{wavecoef} is applied. Bootstrapped lifting trajectories can be used to improve estimation. The authors present a simulation considering small missing data percentages, from 5\% to 20\% using 50 bootstrapped trajectories. More details can be found in \cite{jansen2001scattered} and \cite{knight2017wavelet}.
\subsubsection{Craigmile and Mondal (2020)'s wavelet method}\label{craig}
Let $\{X_t\}_{t\in\Z}$ be a stationary Gaussian long-range dependent time series of interest. The idea behind \cite{craigmile}'s estimator is based on undecimated wavelet analysis of $X_t$ using the Daubechies' class of wavelets, characterized by the filter width $L>2$. A fully detailed explanation of  \cite{craigmile}'s estimator would be lengthly. However, for completeness, a brief description will be provided here. Let $\{h_{j,l}\}_{j,l}$ for $j\in\{0,1,\cdots\}$ and $l\in\{0,\cdots, L-1\}$ be a Daubechies' wavelet filter of even width $L$. The undecimated wavelet representation of
$\{X_t\}_{t\in\Z}$ has coefficients given by
\[W_{j,t}=\sum_{l=0}^{L-1}h_{j,l}X_{t-l}.\]
Since $\{X_t\}_{t\in\Z}$ is a Gaussian process, for each $j$, $\{W_{j,t}\}_{t\in\Z}$ is a zero mean stationary Gaussian process. Let $v_j^2:=\var(W_{j,t})$. \cite{craigmile} show  that, in this case, for $d\in(0,1/2)$,
\begin{equation}\label{wavecoef}
\log(v_j^2)\approx C+(2d-1)j\log(2),
\end{equation}
for large $j$ and some constant $C$. Given an estimator of $v_j^2$, $\hat{v}_j^2$ for $j\in\{j_0,\cdots, j_0+m\}$ for positive $j_0$ and $m$, $d$ can be estimated by  regressing  $\log(\hat{v}_{j_0}^2),\cdots,\log(\hat{v}_{j_0+m}^2)$ in $j_0,\cdots, j_{0+m}$ plus an intercept. The main contribution in \cite{craigmile} is to propose an unbiased, consistent, and asymptotically normally distributed estimator for $v_j^2$ in long-range dependent processes containing missing data. However, to estimate $d$ from  $\hat{v}_j^2$, using a regression approach will depend on an unknown dispersion matrix. The authors propose an approximation for this matrix, leading to an estimator of $d$ called the full estimator (called FULL here). A second estimator is also inspired by the work of \cite{abry1998wavelet} for complete long-range dependent processes, using only the diagonal elements of the full dispersion matrix to estimate $d$. The authors called it the diagonal estimator (called Abry here). We refer the reader to \cite{craigmile} for more details.

\subsection{Traditional methods}\label{trad}

As mentioned in the introduction, there are several estimators to estimate $d$ when the time series is complete. These estimators typically cannot handle missing data naturally. However, they can still be used after the imputation of the missing data. In this work we shall consider five well-known semiparametric estimators for $d$. These are the Rescaled Range Method (R/S) \citep{hurst1951long}, Geweke and Porter-Hudak's (GPH) estimator \citep{geweke1983estimation}, the Local Whittle estimator (LW) \citep{robinson1995gaussian}, the Exact Local Whittle estimator (ELW) \citep{ELW} and the DFA-based estimator (DFA) of \cite{dfa}. More details regarding these estimators can be found in the Supplementary Material.
\subsection{Imputation methods}\label{input}
In this section, we review some of the imputation methods applied in the Monte Carlo simulation and introduce a new one. For a review of some of the \texttt{R} packages available for time series imputation and their performance in the estimation of ARMA models, see \cite{comp1} and \cite{comp2}.

\subsubsection*{Mean Substitution}
The mean substitution is among the simplest imputation methods available. It is based on substituting missing data with the mean calculated over the observed time series values.
\subsubsection*{Linear interpolation}
Interpolation using some simple model is quite a common practice in the literature. This is achieved by a simple linear interpolation in the vicinity of the missing data. If $y_t$ is missing, we apply a simple linear interpolation between the two nearest observed points. Let $y_{t_1}$ and $y_{t_2}$ be the two closest observed points in time satisfying $t_1<t<t_2$. We impute $y_t$ as $y_t = y_{t_1}+\Big(\frac{y_{t_2}-y_{t_1}}{t_2-t_1}\Big)(t-t_1).$
\subsubsection{A new random substitution method}\label{rnd}

Random substitution is an imputation method based on drawing from a predetermined distribution to substitute a given missing value. The most common is to replace a missing value with a random value drawn from a uniform distribution, typically between the minimum and maximum observed values. This simple method has the advantage that no matter the size of a gap, the imputed values will never be equal. However, it tends to inflate the variance of the time series, altering its underlying distribution and affecting its dependence structure.

In what follows, we propose a random substitution method that inherits information regarding the dependence structure on the immediate vicinity of the missing value being imputed. The proposed method is a hybrid of the last observation carried forward method, which consists in substituting each missing value with the most recent observed value, and the random substitution.
Let $tN(\mu,\sigma^2,a,b)$ denote the truncated normal distribution, truncated in the interval $(a,b)$, with mean $\mu\in\R$ and variance $\sigma^2>0$. If $Z\sim tN(\mu,\sigma^2,a,b)$, $Z$ has density
\begin{equation*}%\label{truncnorm}
f(x;\mu,\sigma^2,a,b)= \frac{\phi\left(\frac{x-\mu}{\sigma}\right)}{\sigma\Big[\Phi\big(\frac{b-\mu}{\sigma}\big)-\Phi\left(\frac{a-\mu}{\sigma}\right)\Big]}I(a<x<b),
\end{equation*}
where $\phi$ and $\Phi$ denote the density and distribution of a standard normal distribution, respectively. Let $y_t$, $t>1$ be a missing value to be imputed. We propose imputing $y_t$ by drawing $y_t\sim tN(y_{t-1}, \sigma^2,y_{(1)},y_{(n)})$, where $y_{(1)}=\min\{y_i:i\in M^\complement\}$ and $y_{(n)}=\max\{y_i:i\in M^\complement\}$,  $M:=\{i:y_i \mbox{ is missing}\}$. The variance parameter $\sigma^2$ can be tuned in order to match the variance in the observed time series (see Section \ref{imp}).
\section{Simulation}\label{ssim}

In this section, we present a Monte Carlo simulation study to compare different approaches for estimating the long-range dependence parameter $d$ in ARFIMA processes, considering 35 combinations of estimation and imputation methods. We examine three contexts: (i) estimating $d$ after imputing missing data using complete-data methods; (ii) applying estimators that can directly handle missing data under various missingness scenarios; and (iii) assessing the performance of missing-data estimators when the series is first imputed. While the third approach may seem counterintuitive given the purpose of such estimators, it reflects practical preference for accuracy regardless of methodological derivation.
\subsection{Data generation process (DGP)}
In the Monte Carlo simulation study, we consider Gaussian ARFIMA processes generated using \rr package \texttt{arfima} \citep{arfima}, which, according to the package's documentation, generates ``a sample from a multivariate normal distribution that has a covariance structure defined by the autocovariances generated for given parameters''. The innovation variance was taken as 1. We generate samples of length 2{,}000 and discarded the first 1{,}000 observations as burn-in.

To generate the time series containing missing values, we first generate the complete time series, as described above, and then sample the appropriate number of time indexes for which the data is to be missing from a discrete uniform distribution in the set $\{2,\cdots,999\}$. By design, the first and last values of the time series are never missing. Finally, the observations related to the sampled time indexes  are set to NA. This procedure is repeated for each replication, resulting in a distinct missing data pattern for each generated time series.

\subsection{Implementation details}\label{imp}
In this Section, we shall review some computational details about each estimator's implementation and the imputation method used in the simulations.

\subsubsection*{Intrinsic methods}
The \rr code for the intrinsic Craigmile and Mondal's wavelet method presented in Section \ref{craig} can be found in Craigmile's Github: {\color{blue}\href{https://github.com/petercraigmile/GappyLRD}{github.com/petercraigmile/GappyLRD}}. We have made a few changes to the code to suit our purposes better. We apply Daubechi's D4 wavelet considering levels from 1 to 7. A set of $R$ orthonormal Slepian tapers is used to estimate a specific dispersion matrix in the presence of missing data. The argument $R$ in the code was kept at its default value of 7. More information can be found in \cite{craigmile}.

As for the intrinsic LoMPE method presented in Section \ref{lompe}, it is implemented in the \rr package \texttt{liftLRD} \citep{liftlrd}. We use the bias-corrected estimator available in the package. The long-range dependence parameter is estimated by weighted least squares, using the slope of the log-linear relationship between artificial scales and the log integrals to re-weight the estimates. The slope in the energy-scale relationship is computed using all wavelet levels. As LoMPE is slower than the other methods, we estimate $d$ using only the minimum integral lifting trajectory, since bootstrapping makes the method too slow for simulation purposes.

\subsubsection*{Traditional methods}
The R/S method was implemented by the authors. The estimator is obtained through ordinary least squares using the \rr function \texttt{lm}. Geweke and Porter-Hudak's estimator is implemented in \rr package \texttt{LongMemoryTS} \citep{longmemoryts}. The first $m=\lfloor1+\sqrt{n}\rfloor$  Fourier frequencies were considered in the estimation. The Local Whittle and Exact local Whittle estimators are also implemented in the package \texttt{LongMemoryTS}. In both cases, we considered $m=\lfloor1+\sqrt{n}\rfloor$. In the ELW, the initial value, $Y_0$ is considered known (or estimated). To fulfill this requirement, we consider the time series $\tilde Y_{t-1}:=Y_t-Y_1$ for $t\in\{2,\cdots,n\}$ so that $\tilde Y_0=0$ is used. See Remark 2 in \cite{ELW} for more details and the package \texttt{LongMemoryTS}'s documentation.

The DFA method was implemented by the authors. Function $F^2_{\mathrm{DFA}}$ is calculated using the \rr package \texttt{DCCA} \citep{dcca}, considering non-overlapping windows. The regression step is conducted using ordinary least squares as implemented \rr function \texttt{lm} and considering $h\in\{50,\cdots,100\}$, (i.e., $l=50$ and $s=50$ - see equation (1.8) in the supplementary material).

\subsubsection*{Imputation methods}
The mean substitution method was implemented by the authors. The Linear interpolation method is  implemented in \rr package \texttt{zoo} \citep{zoo} through function \texttt{na.approx}.  The proposed random interpolation method was implemented by the authors. Hyperparameters $a$ and $b$ were taken as the minimum and maximum observed values, respectively. The variance hyperparameter $\sigma$ is user-chosen. The goal in defining $\sigma$ is to use a value that closely approximates the sample standard deviation calculated over the observed values ($S$) for a variety of values of $d$ and missing data proportions. After a pilot simulation study (supplementary material), we found that $\sigma =S/10$ presented good overall performance.  
\subsection{ARFIMA\texorpdfstring{$(0,d,0)$}{(0,d,0)} scenario}\label{0d0}
In this section we study the estimation of the long-range dependence parameter $d$ for Gaussian ARFIMA$(0,d,0)$ processes with $d\in{0.1,0.2,0.3,0.4}$ and missing data proportions ${0.1,0.2,\cdots,0.7}$. The intrinsic estimators from Section \ref{native} are applied to each generated time series with missing values. Parameter $d$ is also estimated from the originally generated series without missing data (the ``original'' series). The incomplete series are then imputed using the three methods described in Section \ref{input}. For each imputation method and missing proportion, $d$ is estimated using the intrinsic estimators and the five estimators from Section \ref{trad}, yielding 35 estimation methods. The experiment is repeated 1{,}000 times. Due to space limitations, we report results only for $d\in{0.1,0.4}$; the remaining cases appear in the supplementary material.
\subsubsection*{Case \texorpdfstring{$d=0.1$}{d=0.1}}

Table \ref{tab:2} reports the simulation results for $d=0.1$. Estimates of $d$ are grouped by time series type: intrinsic (applied to series with missing data) and mean, linear, and random (applied to series imputed by the respective methods). Results are shown for each missing-data proportion, along with the estimates from the original series in column ``0'', repeated across blocks for convenience. The best estimate in each block is shown in bold.
\begin{table}[ht]	
	\caption{Simulation results for the ARFIMA$(0,0.1,0)$ scenario.}\vspace{0.3cm}
\label{tab:2}
\centering
\small
\begin{tabular}{|c|c|c|c|c|c|c|c|c|c|c|}
\hline \hline
\multicolumn{10}{|c|}{$d=0.1$}\\
\hline \hline
\multirow{2}{*}{Type} & \multirow{2}{*}{Estimator} & \multicolumn{8}{|c|}{Missing} \\ \cline{3-10}
& & $0$ &$0.1$ &$0.2$ &$0.3$ & $0.4$ & $0.5$& $0.6$ & $0.7$ \\
\hline\hline
\multirow{3}{*}{Intrinsic} & Full &0.075&0.078&0.085&0.087&0.107&0.112&0.125&0.165 \\
 & Abry & 0.079&0.082&0.083&0.087&0.092&0.100&0.112&0.141 \\
 & LoMPE & 0.063&0.060&0.058&0.056&0.052&0.051&0.048&0.045 \\
\hline\hline
\multirow{10}{*}{Mean} & Full & 0.075&0.068&0.059&0.052&0.043&0.036&0.026&0.019 \\
 & Abry & 0.079&0.072&0.063&0.056&0.047&0.039&0.029&0.022 \\
 & LoMPE & 0.063&0.057&0.051&0.045&0.039&0.032&0.024&0.017 \\
 & DFA & 0.075&0.068&0.060&0.054&0.040&0.036&0.026&0.016 \\
 & GPH & 0.041&0.028&0.017&0.004&-0.012&-0.029&-0.047&-0.075 \\
 & LW & 0.052&0.049&0.045&0.040&0.037&0.033&0.030&0.023 \\
 & ELW & 0.046&0.043&0.039&0.036&0.036&0.043&0.050&0.075 \\
 & RS & 0.145&0.140&0.135&0.130&0.122&0.118&0.109&0.099 \\
\hline\hline
\multirow{10}{*}{Linear} & Full & 0.075&0.132&0.196&0.269&0.348&0.441&0.544&0.676 \\
 & Abry & 0.079&0.130&0.187&0.254&0.328&0.417&0.519&0.652 \\
 & LoMPE & 0.063&0.106&0.155&0.210&0.273&0.350&0.441&0.553 \\
 & DFA & 0.075&0.074&0.076&0.078&0.080&0.092&0.112&0.145 \\
 & GPH & 0.041&0.037&0.034&0.028&0.025&0.023&0.034&0.045 \\
 & LW & 0.052&0.050&0.044&0.040&0.037&0.034&0.033&0.035 \\
 & ELW & 0.046&0.044&0.040&0.037&0.035&0.031&0.033&0.044 \\
 & RS & 0.145&0.152&0.159&0.168&0.176&0.193&0.212&0.238 \\
\hline\hline
\multirow{10}{*}{Random} & Full &  0.075&0.127&0.183&0.244&0.308&0.380&0.453&0.540 \\
 & Abry & 0.079&0.125&0.176&0.232&0.293&0.365&0.440&0.533 \\
 & LoMPE & 0.063&0.102&0.145&0.192&0.243&0.303&0.369&0.447 \\
 & DFA & 0.075&0.070&0.072&0.069&0.072&0.084&0.107&0.138 \\
 & GPH & 0.041&0.036&0.034&0.023&0.018&0.025&0.034&0.041 \\
 & LW & 0.052&0.048&0.042&0.037&0.033&0.031&0.031&0.032 \\
 & ELW & 0.046&0.042&0.039&0.034&0.032&0.029&0.034&0.042 \\
 & RS & 0.145&0.148&0.152&0.156&0.164&0.178&0.197&0.225 \\
\hline \hline
\end{tabular}
\end{table}

A first look at Table \ref{tab:2} reveals that, when there is no missing data, the intrinsic methods produce the best results. Craigmile and Mondal's estimators presented the best results for complete data among the intrinsic estimators. When missing data is taken into account, for all percentages up to 50\% the intrinsic estimators produce the best results overall by a wide margin. The behavior in extreme cases (60\% and 70\%) is less stable.

The effects of mean imputation are pronounced for all estimators and generally worsen as the percentage of missing data increases. Among the imputation methods, the mean performs worst for most missing proportions, except at 70\%, where it yields the best overall result. All methods underestimate $d$, except R/S, which overestimates it. For most estimators, mean imputation substantially degrades the estimates, often producing relative biases above 50\%, particularly for the frequency-domain methods GPH, LW, and ELW. In contrast, R/S behaves oppositely, yielding the best results under mean imputation even as the missing proportion increases.

The best results from the linear and random imputation methods are similar, with a slight advantage for the linear method. For both methods, the intrinsic estimators and R/S greatly overestimate $d$, and their estimates increase with the proportion of missing data, making them unreliable when more than 20\% of the data is missing. With 10\% missing data, the intrinsic estimators perform best overall. For 20\% or more missing data, DFA yields the best results under both imputation methods. The frequency-domain estimators (GPH, LW, and ELW) consistently underestimate $d$ and perform poorly, often with relative bias above 50\%.

% Boxplots of the results are shown in Figure 1 of the supplementary material. Columns correspond to missing-data proportions (10\%, 40\%, and 70\%). The first row displays intrinsic methods applied to series with missing values, while the remaining rows show results for each imputation method. As expected, the variability of the intrinsic estimators increases with the proportion of missing data, while the bias is affected to a lesser extent.

When data imputation is in place, the variability of the estimators does not seem to be significantly impacted by the percentage of missing data. The bias, however, is highly so. The DFA, ELW, and GPH are the methods presenting the highest overall variability. It is also noteworthy that, for 10\% of missing, the imputation method applied makes little difference in the boxplot within each method, but at 40\% and 70\% the imputation method plays an important role.
\subsubsection*{Case \texorpdfstring{$d=0.4$}{d=0.4}}
Table \ref{tab:5} reports the simulation results for $d=0.4$. Among the intrinsic estimators, Craigmile and Mondal's methods consistently outperform the others, maintaining good performance even with 70\% missing data. For the original time series, the GPH estimator performs best, followed by Craigmile and Mondal's estimators. Unlike the case $d=0.1$, where intrinsic methods performed best without missing data, for $d=0.4$ they perform relatively poorly, particularly LoMPE.

The effects of mean imputation are even more severe than for $d=0.1$, with this method performing worst across all missing-data proportions. As the proportion of missing data increases, estimates from all methods deteriorate and uniformly underestimate $d$, although GPH remains the best performer under mean imputation. When the missing proportion reaches 40\% or more, estimates from all methods become uniformly poor and practically unusable.

The linear and random imputation methods affect the estimation in similar fashion, just as in the case of $d=0.1$. For percentages of missing data of 30\% or above, the DFA, LW, ELW, and R/S increasingly degrade as the percentage of missing data increases beyond 30\%, yielding poor estimates. The same happens to GPH, albeit to a lower degree.

%Boxplots of the results are shown in Figure 4 of the supplementary material. The variability of the spectral density–based estimators (GPH, EL, and ELW) is considerably larger than in the case $d=0.1$ for all imputation methods. Overall, the behavior of the estimators is similar for $d=0.1$ and $d=0.4$ when the missing proportion is 10\%. For intrinsic methods, the boxplots for 10\% and 70\% missing data differ little, apart from slightly wider boxes and several outliers for Full. Under linear and random imputation, wavelet-based estimators perform very poorly at 40\% and 70\% missing data.
%
%
\FloatBarrier
\begin{table}[ht]	
	\caption{Simulation results for the ARFIMA$(0,0.4,0)$ scenario.}\label{tab:5}\vspace{0.3cm}
 \centering
 \small
\begin{tabular}{|c|c|c|c|c|c|c|c|c|c|}
\hline \hline
\multicolumn{10}{|c|}{$d=0.4$}\\
\hline \hline
\multirow{2}{*}{Type} & \multirow{2}{*}{Estimator} & \multicolumn{8}{|c|}{Missing} \\ \cline{3-10}
& & $0$ & $0.1$ &$0.2$ &$0.3$ & $0.4$ & $0.5$& $0.6$ & $0.7$ \\
\hline\hline
\multirow{3}{*}{Intrinsic}
 & Full &0.332  &  0.331  &  0.351  &  0.345  &  0.357  &  0.366  &  0.365  &  0.372 \\
 & Abry & 0.344  &  0.349  &  0.351  &  0.354  &  0.356  &  0.359  &  0.360  &  0.367 \\
 & LoMPE & 0.280  &  0.274  &  0.273  &  0.267  &  0.277  &  0.268  &  0.257  &  0.271 \\
\hline\hline
\multirow{10}{*}{Mean}
 & Full & 0.332  &  0.295  &  0.259  &  0.228  &  0.195  &  0.165  &  0.134  &  0.102 \\
 & Abry & 0.344  &  0.308  &  0.273  &  0.242  &  0.209  &  0.178  &  0.145  &  0.111 \\
 & LoMPE & 0.280  &  0.248  &  0.218  &  0.190  &  0.163  &  0.137  &  0.110  &  0.085 \\
 & DFA & 0.366  &  0.354  &  0.340  &  0.324  &  0.303  &  0.280  &  0.251  &  0.214 \\
 & GPH & 0.400  &  0.384  &  0.372  &  0.351  &  0.331  &  0.303  &  0.273  &  0.225 \\
 & LW & 0.321  &  0.314  &  0.306  &  0.296  &  0.284  &  0.268  &  0.250  &  0.220 \\
 & ELW & 0.383  &  0.370  &  0.352  &  0.328  &  0.303  &  0.275  &  0.242  &  0.213 \\
 & RS & 0.343  &  0.333  &  0.322  &  0.308  &  0.292  &  0.274  &  0.249  &  0.220 \\
\hline\hline
\multirow{10}{*}{Linear}
 & Full & 0.332  &  0.376  &  0.424  &  0.480  &  0.544  &  0.617  &  0.702  &  0.810 \\
 & Abry & 0.344  &  0.381  &  0.423  &  0.473  &  0.530  &  0.598  &  0.681  &  0.787 \\
 & LoMPE & 0.280  &  0.314  &  0.353  &  0.397  &  0.447  &  0.511  &  0.587  &  0.682 \\
 & DFA & 0.366  &  0.362  &  0.359  &  0.354  &  0.349  &  0.346  &  0.341  &  0.349 \\
 & GPH & 0.400  &  0.397  &  0.394  &  0.387  &  0.381  &  0.375  &  0.367  &  0.357 \\
 & LW & 0.321  &  0.318  &  0.313  &  0.308  &  0.300  &  0.292  &  0.282  &  0.270 \\
 & ELW & 0.383  &  0.379  &  0.374  &  0.367  &  0.358  &  0.350  &  0.338  &  0.330 \\
 & RS & 0.343  &  0.341  &  0.339  &  0.336  &  0.333  &  0.334  &  0.332  &  0.337 \\
\hline\hline
\multirow{10}{*}{Random}
 & Full &  0.332  &  0.363  &  0.397  &  0.435  &  0.476  &  0.522  &  0.571  &  0.629 \\
 & Abry & 0.344  &  0.370  &  0.398  &  0.431  &  0.470  &  0.514  &  0.564  &  0.626 \\
 & LoMPE & 0.280  &  0.304  &  0.330  &  0.360  &  0.393  &  0.433  &  0.480  &  0.535 \\
 & DFA & 0.366  &  0.358  &  0.354  &  0.341  &  0.334  &  0.327  &  0.318  &  0.317 \\
 & GPH & 0.400  &  0.396  &  0.390  &  0.380  &  0.372  &  0.361  &  0.347  &  0.332 \\
 & LW & 0.321  &  0.315  &  0.309  &  0.299  &  0.289  &  0.278  &  0.262  &  0.243 \\
 & ELW & 0.383  &  0.376  &  0.370  &  0.358  &  0.347  &  0.336  &  0.318  &  0.300 \\
 & RS & 0.343  &  0.336  &  0.330  &  0.323  &  0.317  &  0.314  &  0.311  &  0.316 \\
\hline \hline
\end{tabular}
\end{table}
The DFA, GPH, ELW, and LW methods present a somewhat comparable overall performance, with a slight advantage for the GPH in most cases. The R/S performs stably for all percentages of missing values, especially for the mean imputation case.

\subsection{ARFIMA\texorpdfstring{$(1,d,1)$}{(1,d,1)} scenario}

In this section we consider the estimation of the long-range dependence parameter $d$ for Gaussian ARFIMA$(1,d,1)$ processes with $d\in\{0.1,0.2,0.3,0.4\}$, $\phi=0.5$, $\theta=0.6$, and missing-data proportions $\{0.1,0.2,\cdots,0.7\}$. The same procedure described in Section \ref{0d0} is followed. Tables 
%and figures 
with the results are provided in the supplementary material. The results are very similar to those for the $(0,d,0)$ case, and the same conclusions apply. This similarity is expected because the estimators considered are semiparametric and focus on the long-range dependence structure, largely ignoring short-range components.
\subsection{Time benchmarking}\label{time}
In this section, we compare the computational speed of each estimator considered in the simulations. Besides which estimator is the fastest to compute, there are a few other questions regarding computational speed that are of interest. For instance, does doubling the length of the time series double the time required to estimate $d$? Is the computational time required to estimate $d$ affected by the strength of the dependence? Is the percentage of missing data a factor in estimation times? What about the imputation method applied? In this section, we study these questions through a series of Monte Carlo simulations.

\subsubsection{Setup}
We conduct routines for each estimator involving several subtasks and record the computation time. The routine is divided into two main tasks. The first applies only to intrinsic estimators and consists of estimating $d$ for Gaussian ARFIMA$(0,d,0)$ series with 20\% and 70\% missing data, considering $d\in{0.1,0.4}$ and sample sizes $n\in{1000,2000}$. Each subtask is executed 1{,}000 times per estimator. All time series are generated and prepared beforehand so that the recorded times reflect only the estimation procedures.

The second main task involves all estimators. First, $d$ is estimated from the original time series. The estimators are then applied to series with 20\% and 70\% missing data after imputation using the three methods described in Section \ref{input}. Each exercise is repeated 1{,}000 times, and the computation time is recorded for each subtask and estimator. As before, all time series are generated and prepared beforehand. Simulations were run serially in \texttt{R} version 4.1.3 on a computer with an Intel Core i5 8600k CPU (3.6GHz, factory settings), 16GB RAM, and Windows 10 Pro.
\subsubsection{Results}

The complete results are presented in Tables \ref{tnat} (time series with missing data) and \ref{tclass} (original and imputed series). 

\begin{table}[ht]	
\small
\renewcommand{\arraystretch}{1.3}
\caption{Time spent to complete the simulation task for the intrinsic estimators. Presented is the total time spent, in seconds, performing the respective subtask.}\vspace{0.4cm}	\label{tnat}
\centering
\begin{tabular}{|l|r|r|r|r|r|r|r|r|}
\hline
\multirow{3}{*}{\begin{tabular}{c}Intrinsic\\Estimator\end{tabular}} & \multicolumn{4}{c|}{$n=1{,}000$} & \multicolumn{4}{c|}{$n=2{,}000$}\\
\cline{2-9}
& \multicolumn{2}{c|}{$d=0.1$}& \multicolumn{2}{c|}{$d=0.4$}& \multicolumn{2}{c|}{$d=0.1$}& \multicolumn{2}{c|}{$d=0.4$}\\
\cline{2-9}
&\multicolumn{1}{c|}{20\%} & \multicolumn{1}{c|}{70\%} &\multicolumn{1}{c|}{20\%} & \multicolumn{1}{c|}{70\%} &\multicolumn{1}{c|}{20\%} & \multicolumn{1}{c|}{70\%} &\multicolumn{1}{c|}{20\%} & \multicolumn{1}{c|}{70\%} \\
\hline
Abry  &  392.7  &  392.4  &  388.2  &  388.2  &  1367.6  &  1342.7  &  1334.9  &  1335.2  \\
Full  &  392.8  &  392.4  &  388.2  &  388.3  &  1353.8  &  1341.8  &  1335.1  &  1335.0  \\
LoMPE &  244.8  &   68.8  &  238.1  &   68.1  &   745.6  &   162.4  &   713.2  &  161.3  \\
\hline
\end{tabular}
\end{table}
\FloatBarrier

\begin{table}[ht]	
\renewcommand{\arraystretch}{1.3}
\caption{Time spent (in seconds) to complete the simulation task considering all estimators and imputed/original time series. Presented is the total time spent (in seconds) performing the subtasks. }\vspace{0.4cm}	\label{tclass}
\centering
%\begin{tabular}{|c|c||c|c|c|c|c|c|c|c|c|}
\begin{adjustbox}{max width=\textwidth}
\begin{tabular}{|c|c|c|c||r|r|r|r||r|r|r|r|r|}
\hline
 $n$ & $d$ & \% & input. & Abry & Full & LoMPE & GPH & LW & ELW & DFA & R/S \\
\hline
\hline
 \multirow{14}{*}{\begin{sideways}1{,}000\end{sideways}} &\multirow{7}{*}{$0.1$} &  \multicolumn{2}{c||}{original}
 & 37.1  & 37.1  & 745.6  & 0.09  & 0.14  & 2.63  & 3.36  & 8.97 \\
 \cline{3-11}
 & & \multirow{3}{*}{0.2} & mean
 & 36.9  & 37.1  & 745.9  & 0.11  & 0.16  & 2.81  & 3.26  & 8.94 \\
 & & & linear
 & 37.2  & 37.0  & 745.2  & 0.10  & 0.16  & 2.64  & 3.21  & 8.97 \\
 & & & rand
 & 36.9  & 36.9  & 744.4  & 0.11  & 0.15  & 2.65  & 3.22  & 9.07 \\
  \cline{3-11}
 & & \multirow{3}{*}{0.7} & mean
 & 36.9  & 36.9  & 747.7  & 0.11  & 0.16  & 2.73  & 3.22  & 8.94 \\
 & & & linear
 & 36.9  & 36.9  & 746.8  & 0.10  & 0.15  & 2.57  & 3.22  & 9.04 \\
 & & & rand
 & 36.9  & 36.9  & 782.8  & 0.11  & 0.16  & 2.58  & 3.36  & 9.28 \\
  \cline{2-11}
 & \multirow{7}{*}{$0.4$} & \multicolumn{2}{c||}{original}
 & 36.6  & 36.6  & 728.3  & 0.11  & 0.11  & 2.19  & 3.19  & 9.03 \\
 \cline{3-11}
 & & \multirow{3}{*}{0.2} & mean
 & 36.5  & 36.5  & 733.1  & 0.11  & 0.11  & 2.39  & 3.20  & 8.88 \\
 & & & linear
 & 36.6  & 37.0  & 735.8  & 0.11  & 0.10  & 2.17  & 3.12  & 8.95 \\
 & & & rand
 & 36.5  & 36.5  & 732.4  & 0.10  & 0.11  & 2.20  & 3.18  & 9.00 \\
  \cline{3-11}
 & & \multirow{3}{*}{0.7} & mean
 & 36.5  & 36.4  & 732.2  & 0.11  & 0.11  & 2.65  & 3.19  & 8.87 \\
 & & & linear
 & 36.5  & 36.5  & 733.5  & 0.11  & 0.11  & 2.33  & 3.18  & 8.87 \\
 & & & rand
 & 36.5  & 36.5  & 735.7  & 0.11  & 0.11  & 2.23  & 3.18  & 8.87 \\
  \hline\hline
\multirow{14}{*}{\begin{sideways}2{,}000\end{sideways}} &\multirow{7}{*}{$0.1$} & \multicolumn{2}{c||}{original}
  & 171.1  & 171.1  & 2784.4  & 0.21  & 0.18  & 25.5  & 5.37  & 39.4 \\
  \cline{3-11}
 & & \multirow{3}{*}{0.2} & mean
 & 170.6  & 168.7  & 2808.4  & 0.20  & 0.18  & 25.6  & 5.38  & 38.7 \\
 & & & linear
 & 168.8  & 166.3  & 2633.2  & 0.19  & 0.18  & 25.1  & 5.28  & 38.4 \\
 & & & rand
 & 166.6  & 167.8  & 2636.9  & 0.19  & 0.21  & 25.0  & 5.30  & 38.4 \\
  \cline{3-11}
 & & \multirow{3}{*}{0.7} & mean
 & 166.7  & 167.6  & 2645.8  & 0.18  & 0.22  & 25.9  & 5.28  & 38.4 \\
 & & & linear
 & 166.9  & 166.5  & 2661.0  & 0.21  & 0.22  & 25.1  & 5.65  & 39.7 \\
 & & & rand
 & 170.5  & 169.9  & 2690.8  & 0.19  & 0.26  & 24.4  & 5.37  & 38.4 \\
  \cline{2-11}
 &\multirow{7}{*}{$0.4$} &  \multicolumn{2}{c||}{original}
 & 165.7  & 165.5  & 2638.1  & 0.17  & 0.14  & 20.4  & 5.27  & 38.4 \\
  \cline{3-11}
 & & \multirow{3}{*}{0.2} & mean
 & 165.8  & 165.4  & 2652.5  & 0.20  & 0.14  & 20.9  & 5.26  & 38.5 \\
 & & & linear
 & 165.8  & 165.4  & 2650.5  & 0.20  & 0.15  & 20.6  & 5.27  & 38.5 \\
 & & & rand
 & 165.6  & 165.6  & 2641.4  & 0.19  & 0.15  & 20.4  & 5.25  & 38.4 \\
 \cline{3-11}
 & & \multirow{3}{*}{0.7} & mean
 & 165.7  & 165.8  & 2635.3  & 0.19  & 0.14  & 23.2  & 5.29  & 38.4 \\
 & & & linear
 & 165.6  & 166.0  & 2638.0  & 0.18  & 0.14  & 20.8  & 5.25  & 38.4 \\
 & & & rand
 & 165.3  & 165.7  & 2637.2  & 0.19  & 0.16  & 20.7  & 5.38  & 38.3 \\
\hline
\end{tabular}
\end{adjustbox}
\end{table}
\subsubsection*{Does doubling the time series' length double the time spent in estimating $d$?}
It depends on the estimator and the setup. Considering only intrinsic estimators (Table \ref{tnat}), increasing the sample size from $n=1{,}000$ to $2{,}000$ took, on average, 3.4 times the amount of time to complete the task for \cite{craigmile}'s estimator. LoMPE took, on average, 2.7 times the amount of time.

From Table \ref{tclass}, doubling the sample size from $n=1{,}000$ to $n=2{,}000$ requires less than twice the time to complete the task, on average, for the estimators GPH (1.81), LW (1.34) and DFA (1.65). The R/S and ELW took, on average, 4.3 and 9.3 times the amount of time to complete the task, respectively. Applying the intrinsic estimators to the original and imputed time series (Table \ref{tclass}) and doubling the sample size from $n=1{,}000$ to $n=2{,}000$ on average, produce an overall increase in the time spent to complete the tasks. LoMPE and \cite{craigmile}'s estimator took about 3.6 and 4.5 times the time spent to complete the tasks, on average, respectively.

\subsubsection*{Does the dependence strength affect computational times?}

It depends on the estimator. Completing the full task in the case $d=0.1$  takes on average longer compared to $d=0.4$ for all classical estimators. More precisely, comparing $d=0.1$ versus $d=0.4$, it takes about 42\% longer for the LW, 18\% for the ELW, 2\% for the DFA, 1\% for the R/S, and 0.25\% for the GPH to complete the task. On the other hand, the time spent to complete the task using the intrinsic estimators is not significantly affected by the dependence strength -- the difference in completion time between $d=0.1$ and $d=0.4$, in absolute value, is no more than 2.3\% .

\subsubsection*{Does the percentage of missing data affect computational times?}

It depends on the estimator and whether the time series contains missing data or not. When missing values are considered, performing the tasks is not significantly affected by the percentage of missing values for estimators Full and Abry. For these estimators, when 20\% of the data is missing, performing the task takes no more than 2\% longer when compared to 70\%. However, for LoMPE, performing the task when 20\% of the data is missing takes about 4 times the time spent when 70\% is missing.

Whether the time series is original or imputed, the Whittle estimators LW and ELW are only slightly affected by the percentage of missing values. They require about 6.6\% and 2\% more time, respectively, to complete the task when 70\% of the data are missing compared to 20\%. The other estimators are affected by no more than 1.2\%.

\subsubsection*{Does the imputation method applied affect computational times?}

The time spent completing the task after imputation does not depend on the percentage of missing values prior to imputation nor on the imputation method applied.

\subsubsection*{Which estimator is the fastest to compute?}
It depends on the metrics and scenario. The intrinsic estimators are capable of handling missing data but they are more involved to calculate than the classic estimators. Hence it is expected that the classical estimators can be computed faster.

Table \ref{total} reports the total time each estimator requires to complete the routine, along with the minimum and maximum time spent on a single subtask. The two variants Full and Abry, require nearly identical total times. This is expected, as they differ only in the final step of the algorithm.
\begin{table}[ht]	
\renewcommand{\arraystretch}{1.2}
\caption{Time spent to complete the simulation task considering all estimators and imputed/original time series. Presented is the average time spent (in seconds) over the percentage of missing and imputation methods to perform the subtasks. }\vspace{0.4cm}	\label{total}
\centering
\small
\begin{tabular}{|c|r|r|r|r|r|r|r|r|}
\hline
\multirow{2}{*}{metric}& \multicolumn{8}{c|}{Original/imputed time series}\\
\cline{2-9}
 & \multicolumn{1}{c|}{Abry} & \multicolumn{1}{c|}{Full} & \multicolumn{1}{c|}{LoMPE} & GPH & LW & \multicolumn{1}{c|}{ELW} & \multicolumn{1}{c|}{DFA} & \multicolumn{1}{c|}{R/S} \\
 \hline
total & 2855.4    & 2852.4    & 47743.3    & 4.18    & 4.31    & 358.3    & 119.7    & 665.8   \\
max & 171.1    & 171.1    & 2808.4    & 0.21    & 0.26    & 25.9    & 5.65    & 39.7   \\
min & 36.5    & 36.4    & 728.3    & 0.09    & 0.10    & 2.17    & 3.12    & 8.87   \\
\hline\hline
metric & \multicolumn{8}{c|}{Time series with missing data}\\
\hline
total & 6941.9    & 6927.5    & 2402.3    &     &     &     &     &     \\
max & 1367.6    & 1353.8    & 745.6    &     &     &     &     &     \\
min & 388.2    & 388.2    & 68.1    &     &     &     &     &     \\
\hline\hline
\end{tabular}
\end{table}
\FloatBarrier
If we only consider the total amount of time spent performing the task, when missing data is considered, the fastest estimator is the LoMPE, followed by Full and Abry ones which were the slowest, taking about 2.9 times the time spent by the LoMPE to complete the task. Among the classical estimators, the fastest is the GPH followed closely by the LW. The third fastest is the DFA, but taking about 27.8 times the time to complete the task of the second one. The fastest among the intrinsic when the original/imputed time series is considered are the Full and Abry, with LoMPE being the slowest. The fastest intrinsic estimator (Full) took an astonishing 682 times GPH's total time to complete the task.

However, looking only at totals may not be ideal. For instance, in Table \ref{tnat}, we observe that when $n=1{,}000$ and the percentage of missing is 20\%, the LoMPE is the fastest intrinsic estimator, while when 70\% of the data is missing, the opposite holds. When $n=2{,}000$, the LoMPE is uniformly faster though. Looking at the results presented in Table \ref{tclass}, we observe that the GPH and LW are the fastest estimators by far. For $n=1{,}000$, the GPH is as fast or faster than LW in all but one subtask, while for $n=2{,}000$, the LW is faster in 10 out of 14 subtasks.

A curiosity is that in  \cite{ELW}, the authors claim (page 1891), based on their simulation experience, that the ELW is about 10 times more expensive to compute than the LW. In our simulations we found this number to be about 83 times more expensive, according to Table \ref{total}. Looking at Table \ref{tclass}, we found that the ELW is never less expensive than 16.1 times the LW, with a top value of 166 times. On average, the ELW is about 76.8 times more expensive to compute than the LW. This discrepancy could be due to the efficiency of the implementation applied in the original paper and the one implemented in package \texttt{LongMemoryTS}, used here.
\subsubsection{Convergence}
The classical semiparametric estimators are computationally stable, as is LoMPE. Only the estimators of \cite{craigmile} present computational issues, failing in about 6\% of trials in most scenarios. The problem is more pronounced when $d=0.1$ and the missing proportion exceeds 40\%. For example, when $d=0.1$ and 70\% of the data is missing, the Full variant fails in about one third of the attempts. These failures stem from the wavelet-based construction of the estimator: in sparse settings, insufficient data across wavelet scales can prevent meaningful estimation. 

\FloatBarrier

\section{Conclusion}\label{disc}
In this work, we conducted an extensive Monte Carlo study on the estimation of the long-range dependence parameter $d$ in time series with missing data. We considered estimators designed to handle missing data directly, as well as classical estimators applied after imputation using three different methods. The scenarios included missing proportions from 10\% to 70\%, different sample sizes, and several values of $d$.

The results indicate that mean imputation should be avoided in long-range dependent time series, with linear or random imputation providing better alternatives. When dependence is weak, intrinsic estimators generally perform best, with a slight advantage for Craigmile and Mondal's estimators due to their numerical stability, low bias and variance, and moderate computational cost. If a classical estimator is used, DFA combined with linear or random imputation is the only viable option, as the others perform very poorly.

Under strong long-range dependence, Craigmile and Mondal's estimators applied to the gappy time series present the overall best results. If imputation is to be used, these same estimators paired with random or linear imputation yield the most consistent results.

Our findings show that increasing the sample size has different effects on different estimators; for LoMPE, Full, Abry, R/S and ELW, doubling the sample size from $n=1{,}000$ to $n=2{,}000$ requires more than twice the time to complete the task, on average, while for the others, it requires less. We found that for most estimators, the value of $d$ has a negligible effect on the time required to perform the estimation (exceptions: LW and ELW). The overall fastest estimators are by far the GPH and LW. Among the intrinsic estimator, the LoMPE was the estimator that performed the fastest for $n=2{,}000$. All estimators are very stable with exception of \cite{craigmile}'s, which, due to its particular construction, is prone to fail in very high missing scenarios.

% Finally, we applied the intrinsic estimators (Full and LoMPE) to PM$_{2.5}$ concentration data from seven monitoring stations across North America, with missing data proportions ranging from approximately 4\% to 62\%. The stations encompass diverse environmental contexts, including rural state parks (Viking Lake, Iowa; Jackson, Wyoming; Copan, Oklahoma), urban-industrial sites (Medicine Hat, Alberta; Tlalnepantla, Mexico), a near-road location (Berkeley Aquatic Park, California), and a remote national park (Crater Lake Rim, Oregon). For stations with lower missing proportions (below 30\%), both estimators showed reasonable agreement. For stations with higher missing rates (above 34\%), LoMPE tended to yield noticeably higher estimates than Full, diverging from the otherwise consistent pattern observed across lower missing scenarios. These empirical findings reinforce the practical considerations for environmental monitoring applications, where data gaps are unavoidable and estimation stability is paramount for reliable inference.

\bibliographystyle{apalike}
\bibliography{gladys}

\includepdf[pages=-]{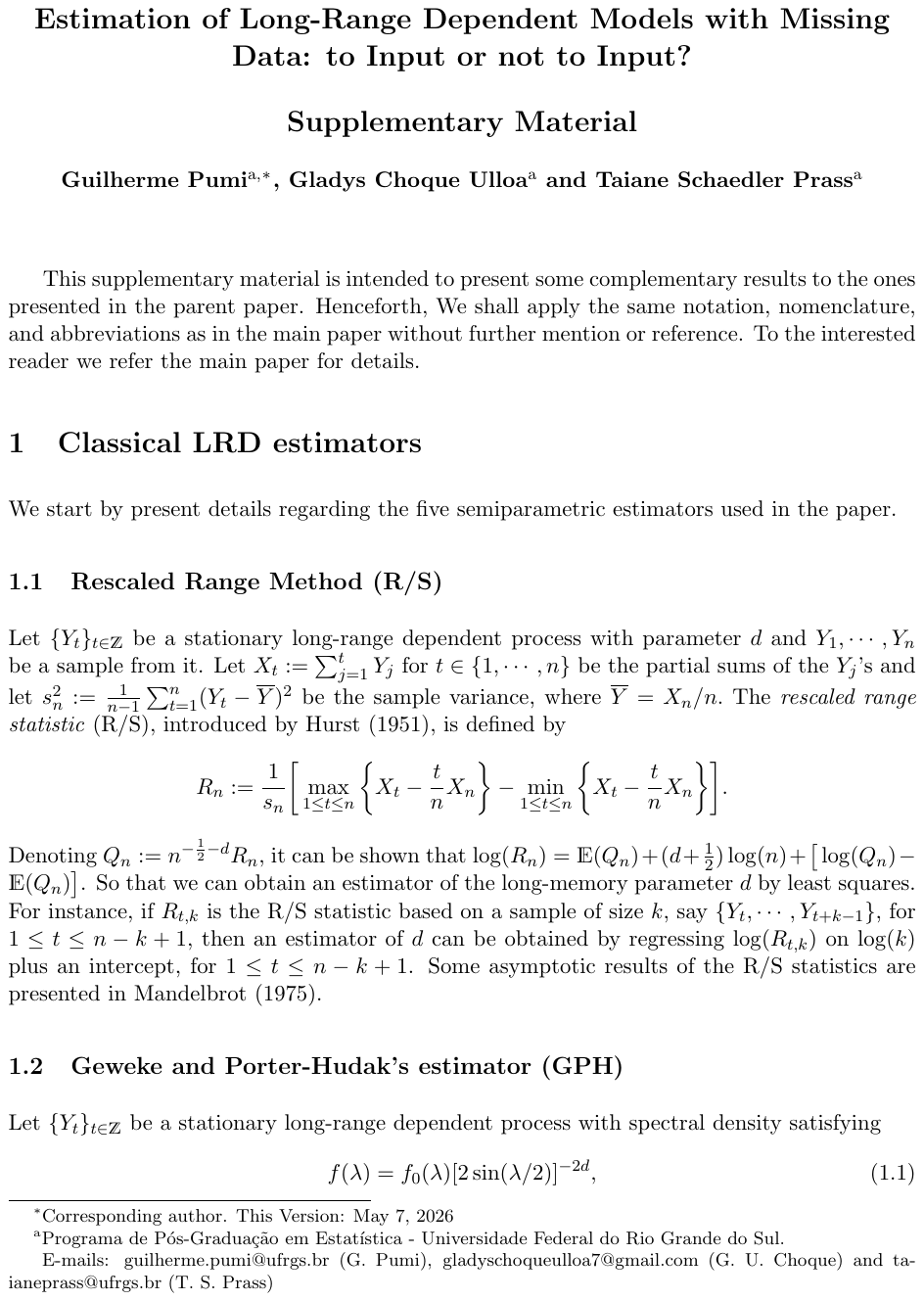}
\end{document}